\documentclass[aps,prb,reprint,amsmath,amssymb,superscriptaddress]{revtex4-1}
\usepackage{textcomp} % degree symbol
\usepackage[pdftex]{graphicx}
\begin{document}

\title{An FPGA-based Instrumentation Platform for use at Deep Cryogenic Temperatures}

\author{I. D. Conway Lamb}
\affiliation{ARC Centre of Excellence for Engineered Quantum Systems, School of Physics, The University of Sydney, Sydney, NSW 2006, Australia}
\affiliation{Microsoft Station Q at Sydney, The University of Sydney, Sydney, NSW 2006, Australia}
\author{J. I. Colless}
\affiliation{ARC Centre of Excellence for Engineered Quantum Systems, School of Physics, The University of Sydney, Sydney, NSW 2006, Australia}
\affiliation{Microsoft Station Q at Sydney, The University of Sydney, Sydney, NSW 2006, Australia}
\author{J. M. Hornibrook}
\affiliation{ARC Centre of Excellence for Engineered Quantum Systems, School of Physics, The University of Sydney, Sydney, NSW 2006, Australia}
\affiliation{Microsoft Station Q at Sydney, The University of Sydney, Sydney, NSW 2006, Australia}
\author{S. J. Pauka}
\affiliation{ARC Centre of Excellence for Engineered Quantum Systems, School of Physics, The University of Sydney, Sydney, NSW 2006, Australia}
\affiliation{Microsoft Station Q at Sydney, The University of Sydney, Sydney, NSW 2006, Australia}
\author{S. J. Waddy}
\affiliation{ARC Centre of Excellence for Engineered Quantum Systems, School of Physics, The University of Sydney, Sydney, NSW 2006, Australia}
\affiliation{Microsoft Station Q at Sydney, The University of Sydney, Sydney, NSW 2006, Australia}
\author{M. K. Frechtling}
\affiliation{Microsoft Station Q at Sydney, The University of Sydney, Sydney, NSW 2006, Australia}
\affiliation{School of Electrical Engineering, The University of Sydney, Sydney, NSW 2006, Australia}
\author{D. J. Reilly$^\dagger$}
\affiliation{ARC Centre of Excellence for Engineered Quantum Systems, School of Physics, The University of Sydney, Sydney, NSW 2006, Australia}
\affiliation{Microsoft Station Q at Sydney, The University of Sydney, Sydney, NSW 2006, Australia}

%\date{\today}

\begin{abstract}
We describe a cryogenic instrumentation platform incorporating commercially-available field-programmable gate arrays (FPGAs) configured to operate well beyond their specified temperature range. The instrument enables signal routing, multiplexing, and complex digital signal processing at temperatures approaching 4 kelvin and in close proximity to cooled devices or detectors within the cryostat. The cryogenic performance of the system is evaluated, including clock speed, error rates, and power consumption. Although constructed for the purpose of controlling and reading out quantum computing devices with low latency, the instrument is generic enough to be of broad use in a range of cryogenic applications.  
\end{abstract}
\maketitle

\begin{figure*}
\includegraphics[width=0.95\textwidth]{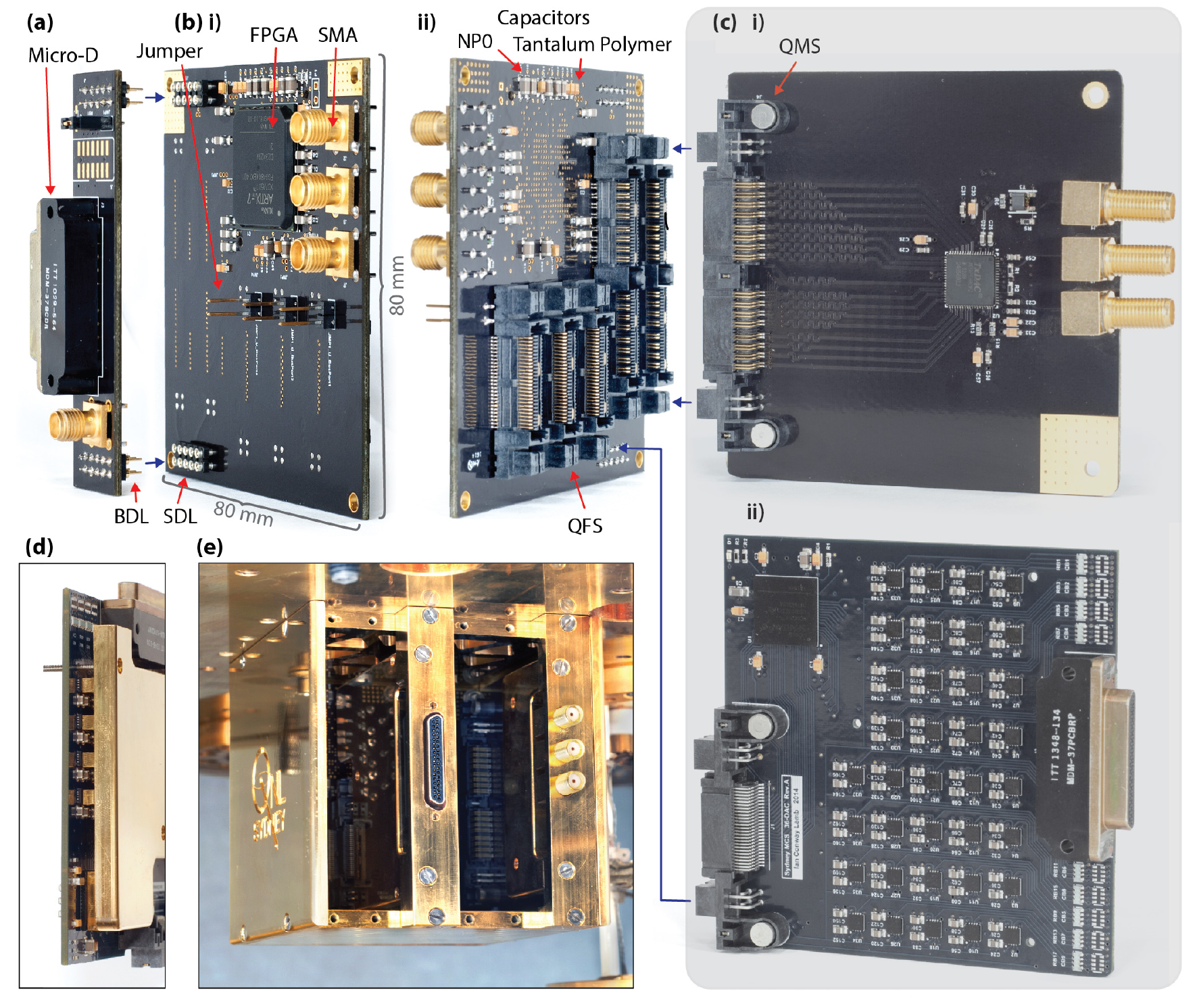}
\caption{Isometric views of modular PCBs. \textbf{(a)} Power-distribution board. \textbf{(b)} (i) Front and (ii) rear view of motherboard with fast (large) and slow (small) expansion connectors visible. \textbf{(c)}(i) High-speed dual-channel DAC daughterboard and (ii) low-speed 36-channel DAC daughterboard. \textbf{(d)} Edge-view of extrusions on the 36-channel DAC heatsink. \textbf{(e)} Modular enclosure, partially filled, and mounted beneath the 4-K stage of a Leiden Cryogenics CS450 dilution refrigerator. A high-speed DAC daughterboard and a low-speed DAC daughterboard fill one high-speed and one low-speed expansion slot, respectively. Details of the components used are given in Table~\ref{tab:BOM}.}
\label{fig:photos}
\end{figure*}

\section{Introduction}
Electronic instrumentation at cryogenic temperatures is widespread in astronomy \cite{SCUBA2}, experimental cosmology \cite{Extreme_electronics,lubin}, and essential to the performance of particle\cite{CUORE,Aprile2012573}, antimatter\cite{1748-0221-10-02-C02023} and single photon\cite{singlephoton} detectors as well as quantum information devices \cite{NatureQIP}. In most configurations, the devices or detectors that require cooling are separated from their room temperature interface and control electronics, typically using low thermal conductivity wiring to cross the often significant thermal gradient. Owing to the Wiedemann-Franz law, thermally resistive wiring must also be electrically lossy, limiting its bandwidth and power carrying capability. For complex instrumentation systems that employ large numbers of wires \cite{CUORE,Trans_edge}, wide bandwidth transmission lines\cite{kinetic_induct_MUX,CollessRSI2}, or low-latency measurement and control, the physical separation between room temperature electronics and the cryogenic device environment poses practical challenges that can impact performance.

Integrating much of the interface electronics inside the high-vacuum stage of the cryostat can partially address these challenges. Embedded cryogenic amplifiers \cite{weinreb} and multiplexing circuits \cite{kinetic_induct_MUX,SCUBA2,Hornibrook_APL,Smith_mux}, for instance, are commonly used to boost weak signals over lengthy transmission lines\cite{Colless_PRL} or to minimise the number of separate cables and feedthrough connectors traversing the vacuum space and temperature gradient. Including in this approach the possibility of operating digital-to-analog converters (DACs)\cite{6908339,Takahashi2014220} and analog-to-digital converters (ADCs) cryogenically\cite{SFQ_ADC,cryoADC}, as well as cryogenic logic and memory systems opens the prospect of digital signal processing, feedback, and realtime control without the need to bring signals up and out of the cryostat. In this configuration the exclusive use of superconducting cables and interconnects also becomes feasible\cite{MCM_ribbonSC,5613217}, greatly reducing the thermal conductivity and cross-section of signal-carrying cables in comparison to lossy normal metals.

Here we describe the design and operation of a modular instrumentation platform for supporting digital signal processing applications at deep cryogenic temperatures, including the functionality of a soft-core processor. Key components of the system are commercially available field-programable gate arrays (FPGAs), configured to function well-beyond their specified operating temperature. The instrument incorporates expansion ports for connection to peripheral data converters such as DACs and ADCs, of use in the autonomous operation and feedback control of quantum information devices \cite{PhysRevApplied.3.024010,Yacobyfeedback,Riste}. Beyond quantum science, the platform is sufficiently generic to be of wide applicability in the read out and control of various cryogenic detectors and devices.

\section{Instrument Design}
\subsection{Overview}
The instrument comprises a motherboard equipped with an Artix-7 FPGA (Xilinx Inc.) and ports for connecting up to five `daughterboards', as shown in Fig. 1. This modular design allows the system to be configured for specific cryogenic applications while providing a generic platform that offers power, digital logic, communication links, and associated thermal management. There are two high-speed and three low-speed connectors for the daughterboard modules on the rear of the motherboard (see Fig.~\ref{fig:photos}(b)(ii)). The two high-speed connectors each have 32 dedicated differential pairs and 8 power pins, and are suitable for high-bandwidth modules including giga-sample per second data converters. The three low-speed connectors share 18 single-ended signalling lines and 8 power pins, have daisy-chained JTAG (IEEE 1149.1) lines, and are suitable for modules which do not need high-bandwidth communication with the motherboard. Various communication protocols are possible using the low-speed connectors; for example, we have made use of a clock-line, a sync-line, and 16 data-lines to transmit 16-bit words using a two-word address/command packet followed by a variable-length data packet. In this paper we do not describe further the separate daughterboard modules, which can be customised for specific applications. 

\subsection{FPGA}
Several semiconductor devices, such as bipolar junction transistors and diodes, suffer from carrier freeze-out at deep cryogenic temperatures, owing to the small fraction of donors that remain ionized. In contrast, the presence of large electric fields in complementary metal-oxide semiconductor (CMOS) devices leads to field-induced donor ionization\cite{Extreme_electronics}. Carrier freeze-out effects can be suppressed by these fields to the extent that digital circuits can continue to operate at deep cryogenic temperatures.  In selecting devices that are compatible with cryogenic operation, the presence of high-K dielectrics to suppress transistor gate-leakage provides an indication that large electric fields are present. The FPGA device central to our instrument, for instance, is manufactured on the TSMC Ltd. 28 nm process node, which makes use of hafnium-oxide between gate and channel\cite{Xilinx:28nmHf}.

The sole active component on the motherboard is a Xilinx Artix-7 FPGA in a 484-pin ball grid array package, and is pin-compatible with 15~k - 100~k (wire-bond package) and 200~k (flip-chip package) logic element versions of the integrated circuit (IC)\cite{Xilinx:A7ProductTable}. The Artix-7 is the lowest power of Xilinx's 7-series FPGAs and has 0.9 to 13.1 Mb of on-chip block random access memory (RAM), and 45 to 740 digital signal processing (DSP) slices, each with a 25~$\times$~18-bit multiplier. DSP configuration options include pre- and post-adders and a 48-bit accumulator~\cite{Xilinx:DSP48E1}. High speed grade (-3) and low power (-2L, -1L) variants are available~\cite{Xilinx:Switching}. The specific device used in our instrument is an XC7A50T-2FGG484C, which has 50~k logic elements, has 2.7 Mb of block RAM and 120 DSP slices, and comes from the common `-2' speed-grade bin in the commercial temperature range (0-85\textdegree C). We also make use of Spartan-3 and Spartan-6 devices, which are configured to be operational at deep cryogenic temperatures. These simpler FPGAs are ideal for adding functionality to the daughterboards, for example, in parsing data to a DAC. 

\subsection{Printed Circuit Board Design}
\begin{figure}
\includegraphics[width=0.48\textwidth]{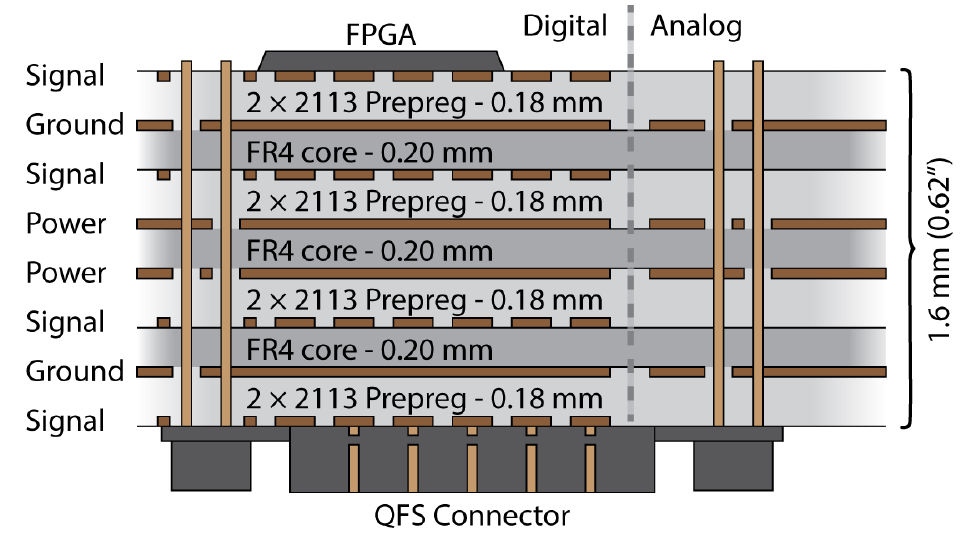}
\caption{Motherboard PCB 8-layer stackup: 35~\textmu m (1~oz) copper layers, with 3 FR4 cores, and 4 prepreg layers. The digital (signal and power) and analog (power only) domains are spatially separated on the PCB, and can have their grounds tied on the power distribution board, outside the fridge, or provided independently. Power pins of the QFS connectors are through-hole pins, and are connected to the power planes, while signal pins are surface-mount. PCB vias are not shown.}
\label{fig:stackup}
\end{figure}

The motherboard is an 80~mm~$\times$~80~mm, 8-layer FR4 printed circuit board (PCB), with all layers using 35~\textmu m (1~oz) of copper, as shown in the stack-up illustrated in Fig.~\ref{fig:stackup}. The board is finished with electroless nickel immersion gold (ENIG) plating. A simplified bill of materials is provided in Table~\ref{tab:BOM}, listing the manufacturers and part numbers of the components.

Capacitors used  on the motherboard and daughterboards for decoupling and filtering are a mix of NP0 ceramic and tantalum polymer (TP). At cryogenic temperatures NP0 capacitors lose negligible capacitance, and maintain low equivalent series resistance (ESR)\cite{Teyssandier:2010,NASA:Capacitors}, but the capacitance density is small. TP capacitors also retain stable ESR but suffer from a reduction in capacitance when cooled\cite{NASA:Capacitors}. Their higher capacitance density however, makes them more suitable when large capacitance is required. Thin-film resistors are used for their temperature stability, as opposed to thick-film resistors, which have been observed to vary dramatically in resistance when cooled. 

\begin{figure*}
\includegraphics[width=0.95\textwidth]{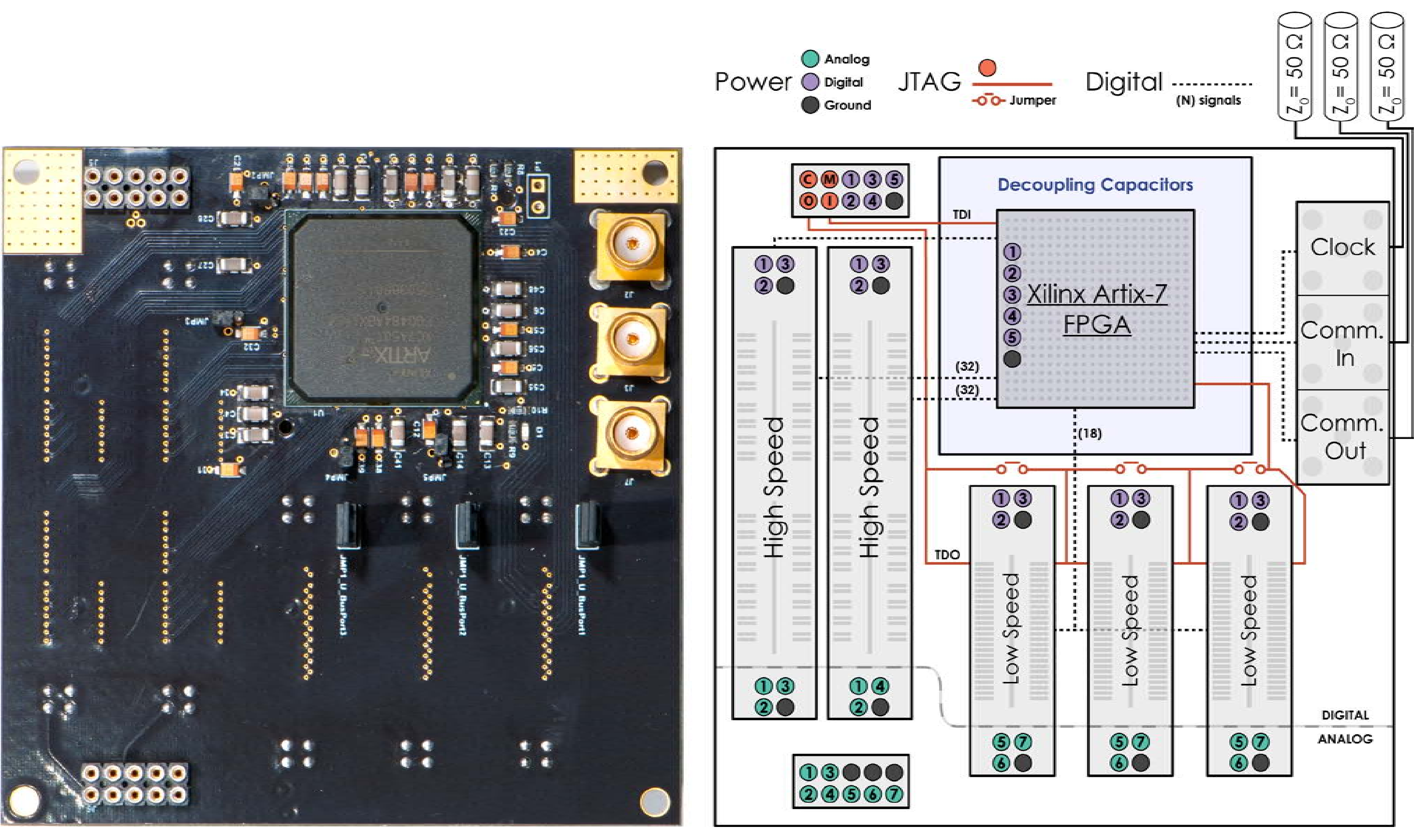}
\caption{Layout of the modular instrumentation platform motherboard. The purple and green circles show distribution of the 5 digital and 7 analog voltage rails, from the 10-pin SAMTEC SDL connectors to the modular expansion connectors (the digital and analog domains are spatially separated, as shown by the dashed line). Red lines indicate the JTAG data signal daisy-chain, with optional jumpers to bypass unused ports. Short dashed lines represent digital signals to and from the FPGA, with numbers indicating the total number of connections.}
\label{fig:diagram}
\end{figure*}
\begin{table}[h]
\caption{Simplified Bill of Materials}
\begin{tabular}{ l l l }
  \hline
Component 				& Manufacturer 	& Part \\
  \hline
FPGA 						& Xilinx			& XC7A50T-2FGG484C \\
Capacitors	 			& AVX 			& TCJ Series (TP)\\
Capacitors		 		& Kemet  			& C Series (NP0)\\
Resistors					& Panasonic 		& ERA-3A Series (thin film)\\
\hline
Connectors				&					&\\
%\hline
High-speed	 			& SAMTEC 		& QFS-032-04.25-L-D-DP-PC4 \\
High-speed$^1$			& SAMTEC 		& QMS-032-01-L-D-DP-RA-PC4 \\
Low-speed		 		& SAMTEC 		& QFS-026-04.25-L-D-PC4 \\
Low-speed$^1$			& SAMTEC 		& QMS-026-01-L-D-RA-PC4 \\
Power						& SAMTEC 		& SDL-105-T-10 \\
Power$^2$				& SAMTEC		& BDL-105-G-E \\
SMA					 	& Cinch 			& 142-0701-211 (vertical) \\
SMA$^1$					& Cinch	 		& 142-0701-551 (right angle) \\
Micro-D$^{1}$			& Glenair 			& MWDM5L-37PCBR-.110 \\
Micro-D$^{2}$			& Glenair 			& MWDM5L-37SCBR-.110 \\
  \hline
\end{tabular}
\label{tab:BOM}
$^1$Daughterboard component, $^2$ Powerboard component
\end{table}

\subsection{Communication and Clocking}
Communication between the cryogenically operated FPGA and room temperature instruments is provided via stainless steel coaxial cables that mate with the SMA connectors, with optional 50~$\Omega$ termination, on the motherboard. A global clock signal is also provided to the instrument in this way. Although alternative clocking and communication protocols are possible, our typical configuration brings three coaxial cables into the cryostat for the FPGA clock, for a serial input signal, and for a serial output. The clock is used to generate various other internal clocks required: for operation of the serial interface, clocking data for the low-speed connectors, and to run internal logic. 

\subsection{Power Supply and Programming}
The power supply for the instrument as well as the FPGA programming signals are carried from room temperature using beryllium-copper cryogenic loom wire to the 37-pin Micro-D connector on the power-distribution module (see Fig. 1(a)). The module distributes power and programming signals via two 10-pin sockets on the front of the motherboard; one socket for digital power and programming, and one socket for analog power. Five digital and seven analog voltage lines, plus analog and digital ground lines, are supplied. Four-terminal sensing, with force and sense pairs tied at the power distribution board, compensates for loom wire resistance ensuring correct voltages at the instrument.

Programming and debugging of the motherboard and daughterboards is performed using a Xilinx Platform Cable USB II. The cable uses a standard 4-pin interface: the clock (TCK) and mode-select (TMS) signals are shared between the motherboard FPGA and the low-speed connectors; the test-data-in (TDI) and test-data-out (TDO) signals are daisy-chained from the motherboard FPGA to each of the low-speed connectors. In case a daughterboard module is not installed in a low-speed connector slot, a jumper is provided on the front of the motherboard to connect the unused TDI and TDO, ensuring a complete JTAG chain (see Fig.~\ref{fig:diagram}).

\subsection{Thermal Management}
The entire instrument comprising power-distribution module, motherboard, and daughterboard modules is housed in a gold-plated copper chassis, which ensures good thermal coupling between the active electronic components and cryostat. The overall length and width of the instrument is 96.5~mm and 88~mm, and the height is 92~mm. The motherboard and each daughterboard have their own copper mounts which feature extrusions to make direct thermal contact to the packaging of the integrated circuits. In Fig.~\ref{fig:photos}(d), extrusions are shown which thermally connect individual DAC ICs on a low-speed multi-channel DAC daughterboard module. The instrument is installed at the 4-K stage of a cryogen-free dilution refrigerator, as shown in Fig.~\ref{fig:photos}(e). Unused slots can be covered with blank copper panels to reduce electromagnetic interference. The instrument is cooled slowly from room temperature in the presence of helium exchange gas, which is evacuated when a temperature of  4~K is reached.

\section{FPGA Operation}
Modern FPGAs have complex internal architectures with many subsystems for specialised tasks. The reconfigurable general-purpose digital logic comprises a large number of configurable logic blocks (CLBs). A switch matrix for each CLB connects it to the general routing matrix. Each CLB contains flip-flops (FFs), look-up tables (LUTs), multiplexers, basic logic, and memory\cite{Xilinx:7CLB}. FPGAs incorporate DSP slices which contain hardware multipliers and accumulators, for specialised high-throughput operations. In addition, input and ouput (IO) buffers can be configured to suit various single-ended and differential voltage specifications. A summary of the operation of these components at cryogenic temperatures is given in Table~\ref{tab:summary}, for Artix-7 and Spartan FPGAs.
\begin{table}[h]
\caption{Cryogenic Operation of FPGA}
\begin{tabular}{ l l}
\hline
Single-ended IO & Operational \\
Differential inputs & Operational \\
Differential outputs~~~~~~& Spartan-3 only \\
PLLs & Non-operational$^1$ \\
Digital logic\quad & Operational \\
Block RAM & Operational$^2$ \\
DSP slices & Operational$^2$ \\
\hline
\end{tabular}
\\$^1$Tested on Motherboard using Artix-7 
\\$^2$Tested on Motherboard and  Daughterboards with Artix-7 and Spartan-6 \label{tab:summary}
\end{table}

\subsection{IO Voltage Characterisation}

\begin{table*}[ht]

\caption{Instrument IO Parameters}

\begin{tabular}{ l r r r r r r }

\hline

& \multicolumn{2}{c}{~Artix-7} & \multicolumn{2}{c}{~Spartan-6$\,\,\,$}
& \multicolumn{2}{c}{~Spartan-3$\,\,\,$} \\

& ~~300~K & ~~4~K & ~~300~K & ~~4~K$\,\,\,$ & ~300~K & ~~4~K$\,\,\,$ \\

\hline

V$_\text{IH}$ single-ended LVCMOS (V) & 1.16 & 1.22 & 2.39 &
2.51$\,\,\,$ & 1.50 & 1.61$\,\,\,$ \\

V$_\text{IL}$ single-ended LVCMOS (V) & 1.09 & 1.11 & 2.14 &
2.24$\,\,\,$ & 1.42 & 1.47$\,\,\,$ \\

V$_\text{IH}$ differential LVDS (mV) & 5 & 18 & 11 & 11$\,\,\,$ & 18 &
13$\,\,\,$ \\

V$_\text{IL}$ differential LVDS (mV) & -39 & -55 & 18 & -33$\,\,\,$ &
-39 & -35$\,\,\,$ \\

Pull-up resistance (k$\Omega$) & 20 & 17 & 10 & 7.7$\,\,\,$ & 10 &
6.6$\,\,\,$ \\

Differential resistance ($\Omega$) & 96 & 86 & 101 & 93$\,\,\,$ & 108 &
105$\,\,\,$ \\

Differential output voltage (mV) & 435 & 42 & 372 & 1569$^*$ & 387 &
582$^*$ \\

Common-mode differential output voltage (mV) & 1120 & 227 & 1242 &
1202$^*$ & 1056 & 1652$^*$ \\

\hline
\end{tabular}
\label{tab:IO}
\begin{tabular}{p{\textwidth}}
\\
{Instrument input and output logic thresholds were measured for LVCMOS and LVDS
standards, for the Artix-7, Spartan-6 and Spartan-3, at a temperature of 300~K and 4~K.
The following characteristics are tabulated: input-high (V$_\text{IH}$) and
input-low (V$_\text{IL}$) thresholds, for 3.3~V LVCMOS signals; the
differential input thresholds for LVDS signals with a 1.2~V common-mode
voltage; the resistance of an internal pull-up resistor for 3.3~V
LVCMOS, when the input is held at 0~V; the dc differential input
resistance of an internal differential termination, for a 400~mV
differential signal and 1.2~V common-mode voltage; and the differential
and common-mode output voltages of LVDS signals, measured using an
oscilloscope with a 100~$\Omega$ terminating resistor at room temperature. Asterisk indicates an average reading for voltages that fluctuate in time.}
\end{tabular}
\end{table*}
We first investigate if cooling the instrument leads to variations in the FPGA switching voltage levels associated with the single-ended low-voltage CMOS (LVCMOS) and low-voltage differential-signalling (LVDS) logic standards. Input thresholds are measured by applying a dc input voltage and measuring the minimum voltage which always gives a high output (V$_\text{IH}$) and the maximum voltage which always gives a low output (V$_\text{IL}$). A common-mode voltage of 1.2~V is used for LVDS. Results are presented in Table~\ref{tab:IO}. Both LVCMOS and LVDS input thresholds change negligibly with cooling to cryogenic temperatures allowing standard operation. We observe a decrease in the resistance of pull-up resistors and differential termination resistors with cooling, but note that these variations can be compensated for with careful circuit design. LVCMOS outputs function normally.

The parameter that varies the most with cryogenic operation is the LVDS output voltage. Below 50~K, the common-mode and differential output voltages both decrease dramatically on the Artix-7, and both increase dramatically on the Spartan-6. Only the Spartan-3 FPGA exhibits functioning differential output signalling. The operation of LVDS outputs are likely linked to internal bandgap voltage reference offsets occurring at low temperature.
\begin{figure}
\includegraphics[width=0.48\textwidth]{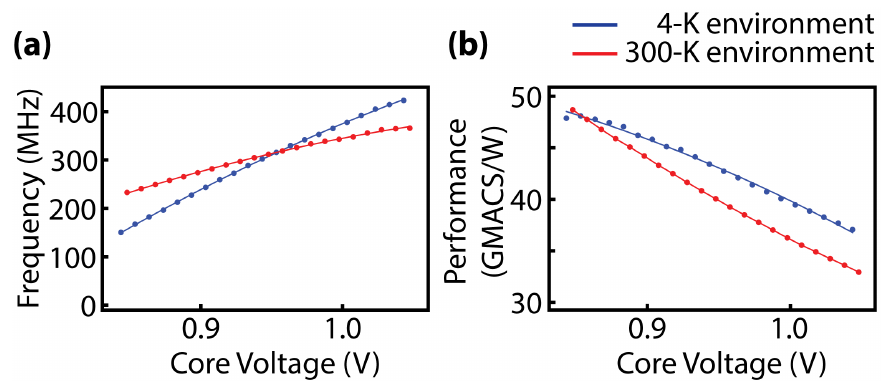}
\caption{\textbf{(a)} Maximum operating frequency of the instrument with the Artix-7 running 30~MAC blocks. \textbf{(b)} Performance in GMACS/W (Giga-MACs per second per watt) at maximum operating frequency. Blue-coloured data are taken with the FPGA in the instrument mounted at the 4-K stage of the fridge with red-coloured data indicating room temperature operation. The solid lines are guides to the eye.}
\label{fig:fmax}
\end{figure}
\subsection{Performance and Soft Processor Operation}
To demonstrate the functionality of the instrument's general-purpose digital logic at deep cryogenic temperatures, we have implemented an embedded soft processor, a Xilinx `MicroBlaze'\cite{Xilinx:Microblaze}, executing standard C-code, and loaded onto the FPGA via the JTAG interface. We have run implementations with logic utilisation of up to 8569 FFs (13\% utilisation), 7190 LUTs (22\% utilisation), and 1476~kb of block RAM (55\% utilisation). Based on this demonstration we expect that similar IP-cores, for instance the Cortex-M1 implementations available from ARM Ltd.\cite{ARM_M1}, could be embedded in our instrument and run in the cryogenic environment.

Since the instrument generates heat in proportion to its clock speed, performance is constrained by the available cooling power of the cryostat for a given temperature. With our instrument mounted at the 4-K stage of a standard cryogen-free dilution refrigerator however, we find that significant computational performance is possible without adversely affecting the mixing chamber base temperature. To benchmark  performance, we carry out a series of tests involving multiply-and-accumulate (MAC) blocks, noting that these MAC operations form the basis of many DSP applications including filters, window functions, down-converters and Fourier transforms. The setup for our test comprises the execution of 30 DSP48E1 slices\cite{Xilinx:DSP48E1}, configured as 16$\times$16-bit MAC operations and clocked with an external (room temperature) variable source from 0 - $\sim$400 MHz. We proceed by comparing the output of 1000 accumulated multiplications of pre-generated random numbers to the expected result, with error rates recorded. The test is performed as a function of core voltage, comparing instrument behaviour at room temperature and 4~K. 

The maximum operating frequency of the instrument is determined as the FPGA clock frequency at which there are no errors over 32 repeat test runs, corresponding to a total of 960,000 MAC operations. The maximum clock frequency is a function of both the core voltage and temperature of the FPGA, as shown in Fig.~\ref{fig:fmax}(a). We note that slightly higher frequency clocking is possible at cryogenic temperatures, when operating at the nominal core voltage of 1.0~V.
\begin{figure}
\includegraphics[width=0.47\textwidth]{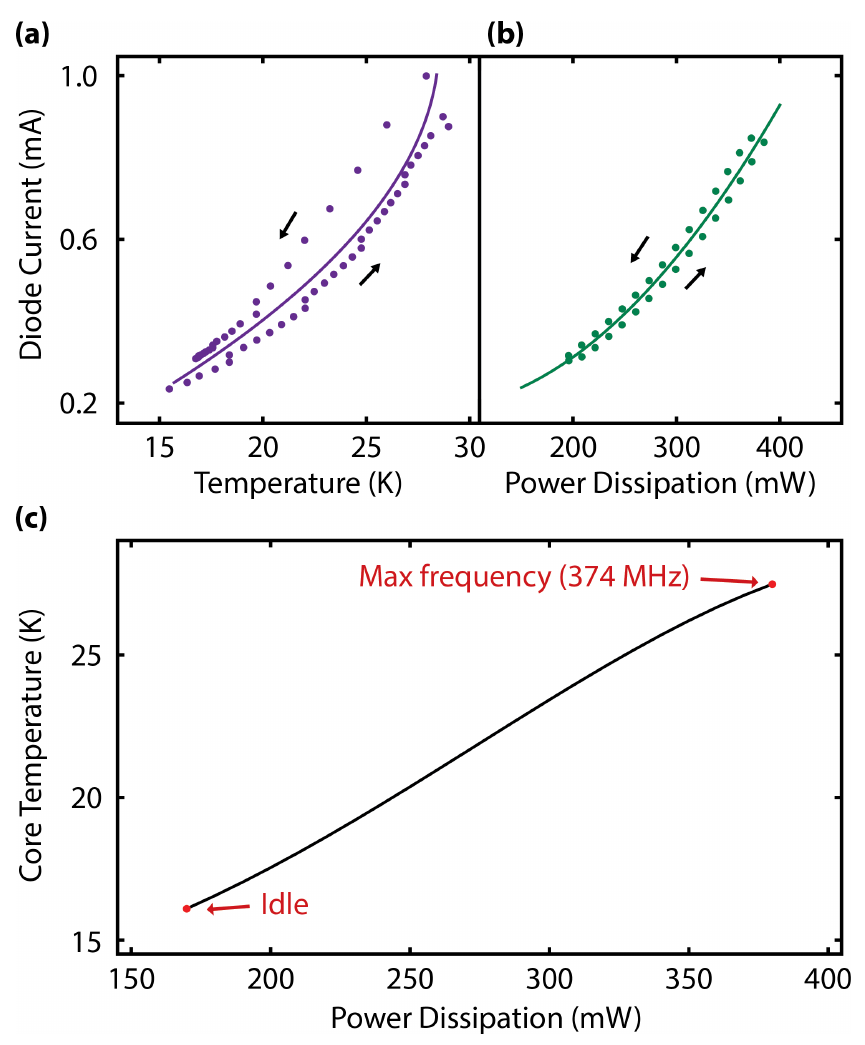}
\caption{Estimating the temperature of the Artix-7 FPGA die as a function of power dissipated. With the FPGA off, (a) shows the diode current in response to a voltage, measured as the refrigerator temperature is increased and decreased to provide a calibration (arrows show direction of temperature sweep). As some hysteresis in the diode response is observed with heating and cooling, our calibration is taken from a line-of-best fit. In (b) the diode current is monitored as a function of FPGA power, allowing a functional dependence of core temperature on power to be determined. This function is plotted in (c), extrapolating to conditions at maximum operating frequency and idle (only static power).}
\label{fig:core_temp}
\end{figure}

\subsection{Power Dissipation and Operating Temperature}
We have compared the performance and power dissipation of the motherboard Artix-7 FPGA when operating at room temperature and inside the dilution refrigerator, as shown in Table~\ref{tab:performance} and Fig.~\ref{fig:fmax}(b). Cooling the instrument to 4 K increases the static power but decreases the dynamic power. In the case of the dynamic power, we evaluate power dissipation using only the core voltage data from the DSP-block test, expressed as power per clock-rate (mW/MHz) or equivalently, average energy per MAC operation (nJ/MAC).

\begin{table}[h]
\caption{Artix-7 power and performance}
\begin{tabular}{ l r r }
  \hline
  & ~300~K & ~~~4~K  \\
  \hline
Static power (mW) & 95 & 170 \\
Dynamic power (mW/MHz  or  nJ/MAC) & 22.9 & 21.5 \\
Maximum frequency at 1.0~V (MHz) & 344 & 374 \\
  \hline
\end{tabular}
\label{tab:performance}
\end{table}

Although we can ensure that the outer casing of the FPGA is well-thermalised to the 4 kelvin stage of the dilution refrigerator, it is likely that the die exhibits hot-spots and an overall elevated temperature with respect to its packaging. We provide a course measure of the die temperature by directly accessing a semiconductor diode via external pins of the FPGA, performing a calibration of the resistance of the diode as a function of the refrigerator temperature with the FPGA unpowered (see Fig.5 (a)). During the DSP-block test, the diode current was recorded as a function of clock frequency to determine the relationship between the on-chip temperature ((Fig. 5(b)) and the power dissipation, as shown in Fig. 5(c). We estimate that the core temperature rises from 16~K at 170~mW (idle) to 27.5~K at 380~mW. We thus determine an approximate thermal resistance between the chip and the 4-K stage of the refrigerator to be 55~K/W. Despite the FPGA die having an elevated internal temperature, the instrument and its connectors remain in close thermal contact with the 4-K stage, even at the highest clock rates. 

\section{Discussion}
Integrating FPGA-based instrumentation directly in the cryogenic environment with cooled devices or detectors has technical advantages such as the use of miniaturised, high-density superconducting cabling, and in enabling the operation of cryogenic multiplexers, DACs and ADCs. In addition to these practical aspects, the functionality of FPGAs at cryogenic temperatures provides a path to establishing complete instrumentation solutions that take advantage of reduced temperatures to improve performance. Cooling analog circuits, for example, leads to a reduction in thermal noise\cite{CryoSiGeAmp} and an increase in the transconductance and gain of transistors \cite{Cressler:2010}. For digital systems, lower temperatures improve carrier mobility, reduce the interconnect resistance, and lower the subthreshold swing, leading to higher clock speeds and lower power dissipation\cite{Hanamura:1986,Clark:1992}. Integrated with cryogenic FPGA, it is anticipated then that these improvements to semiconductor-based circuits can lead to enhanced performance of the classical control and readout hardware needed to scale-up quantum computing devices. Further improvements are likely found via the use of structured-ASICs (application-specific integrated circuits), that hard-wire program implementations during fabrication. Such devices can lead to significant reductions in power dissipation at cryogenic temperatures. 

We also draw attention to the potential use of cryogenic FPGA in the context of superconducting digital electronics, interfacing with rapid single flux quantum (RSFQ)\cite{likharev1991rsfq} or reciprocal quantum logic (RQL) devices\cite{rql}. Cryogenic systems featuring such devices are already commercially available\cite{vernik2007cryocooled}, controlled via room temperature FPGAs. Operating the control electronics cryogenically would enable more tightly-integrated and compact instruments of potential use in systems requiring low-latency, high-speed feedback control.

\section{Conclusion}
We have developed a cryogenic FPGA-based modular instrumentation platform for the readout and control of detectors and devices at temperatures approaching 4~K. Our instrument makes use of three models from Xilinx Inc. (Artix-7, Spartan-6, and Spartan-3), which have been shown to operate in the cryogenic, high-vacuum environment of a dilution refrigerator. It is possible to use much of the functionality of the more powerful, Artix-7-FPGA, including the operation of a soft-core processor and DSP blocks. We anticipate that such FPGA-based instruments can enhance the performance of the classical control hardware needed in the operation of next-generation quantum technologies.

\section{Acknowledgements}
We thank D. Johnson for useful conversations. This work was supported by Microsoft Research, the Office of the Director of National Intelligence, Intelligence Advanced Research Projects Activity (IARPA), through the Army Research Office grant W911NF-12-1-0354, the Army Research Office grant W911NF-14-1-0097, and the Australian Research Council Centre of Excellence Scheme (Grant No. EQuS CE110001013).

$\dagger$ Corresponding author, email: david.reilly@sydney.edu.au 

%\bibliography{cryo_fpga_bib}

\begin{thebibliography}{41}%
\makeatletter
\providecommand \@ifxundefined [1]{%
 \@ifx{#1\undefined}
}%
\providecommand \@ifnum [1]{%
 \ifnum #1\expandafter \@firstoftwo
 \else \expandafter \@secondoftwo
 \fi
}%
\providecommand \@ifx [1]{%
 \ifx #1\expandafter \@firstoftwo
 \else \expandafter \@secondoftwo
 \fi
}%
\providecommand \natexlab [1]{#1}%
\providecommand \enquote  [1]{``#1''}%
\providecommand \bibnamefont  [1]{#1}%
\providecommand \bibfnamefont [1]{#1}%
\providecommand \citenamefont [1]{#1}%
\providecommand \href@noop [0]{\@secondoftwo}%
\providecommand \href [0]{\begingroup \@sanitize@url \@href}%
\providecommand \@href[1]{\@@startlink{#1}\@@href}%
\providecommand \@@href[1]{\endgroup#1\@@endlink}%
\providecommand \@sanitize@url [0]{\catcode `\\12\catcode `\$12\catcode
  `\&12\catcode `\#12\catcode `\^12\catcode `\_12\catcode `\%12\relax}%
\providecommand \@@startlink[1]{}%
\providecommand \@@endlink[0]{}%
\providecommand \url  [0]{\begingroup\@sanitize@url \@url }%
\providecommand \@url [1]{\endgroup\@href {#1}{\urlprefix }}%
\providecommand \urlprefix  [0]{URL }%
\providecommand \Eprint [0]{\href }%
\providecommand \doibase [0]{http://dx.doi.org/}%
\providecommand \selectlanguage [0]{\@gobble}%
\providecommand \bibinfo  [0]{\@secondoftwo}%
\providecommand \bibfield  [0]{\@secondoftwo}%
\providecommand \translation [1]{[#1]}%
\providecommand \BibitemOpen [0]{}%
\providecommand \bibitemStop [0]{}%
\providecommand \bibitemNoStop [0]{.\EOS\space}%
\providecommand \EOS [0]{\spacefactor3000\relax}%
\providecommand \BibitemShut  [1]{\csname bibitem#1\endcsname}%
\let\auto@bib@innerbib\@empty
%</preamble>
\bibitem [{\citenamefont {Holland}\ \emph {et~al.}(2003)\citenamefont
  {Holland}, \citenamefont {Duncan}, \citenamefont {Kelly}, \citenamefont
  {Irwin}, \citenamefont {Walton}, \citenamefont {Ade},\ and\ \citenamefont
  {Robson}}]{SCUBA2}%
  \BibitemOpen
  \bibfield  {author} {\bibinfo {author} {\bibfnamefont {W.}~\bibnamefont
  {Holland}}, \bibinfo {author} {\bibfnamefont {W.}~\bibnamefont {Duncan}},
  \bibinfo {author} {\bibfnamefont {B.}~\bibnamefont {Kelly}}, \bibinfo
  {author} {\bibfnamefont {K.}~\bibnamefont {Irwin}}, \bibinfo {author}
  {\bibfnamefont {A.}~\bibnamefont {Walton}}, \bibinfo {author} {\bibfnamefont
  {P.}~\bibnamefont {Ade}}, \ and\ \bibinfo {author} {\bibfnamefont
  {E.}~\bibnamefont {Robson}},\ }\href@noop {} {\bibfield  {journal} {\bibinfo
  {journal} {Proc. SPIE: Millimeter and submillimeter detectors for astronomy.
  T.G. Philips and J. Zmuidzinas (eds.) \
  }\textbf {\bibinfo {volume} {4855}},\ \bibinfo {pages} {1} (\bibinfo {year}
  {2003}). (arxiv.org/pdf/astro-ph/0606338)}}\BibitemShut {NoStop}%
\bibitem [{\citenamefont {Cressler}\ and\ \citenamefont
  {Mantooth}(2013)}]{Extreme_electronics}%
  \BibitemOpen
  \bibfield  {author} {\bibinfo {author} {\bibfnamefont {J.~D.}\ \bibnamefont
  {Cressler}}\ and\ \bibinfo {author} {\bibfnamefont {H.~A.}\ \bibnamefont
  {Mantooth}},\ }\href@noop {} {\emph {\bibinfo {title} {{Extreme Environment
  Electronics, (CRC Press)}}}}\ (\bibinfo {year} {2013})\BibitemShut {NoStop}%
\bibitem [{\citenamefont {Kogut}\ \emph {et~al.}(2004)\citenamefont {Kogut},
  \citenamefont {Wollack}, \citenamefont {Fixsen}, \citenamefont {Limon},
  \citenamefont {Mirel}, \citenamefont {Levin}, \citenamefont {Seiffert},\ and\
  \citenamefont {Lubin}}]{lubin}%
  \BibitemOpen
  \bibfield  {author} {\bibinfo {author} {\bibfnamefont {A.}~\bibnamefont
  {Kogut}}, \bibinfo {author} {\bibfnamefont {E.}~\bibnamefont {Wollack}},
  \bibinfo {author} {\bibfnamefont {D.~J.}\ \bibnamefont {Fixsen}}, \bibinfo
  {author} {\bibfnamefont {M.}~\bibnamefont {Limon}}, \bibinfo {author}
  {\bibfnamefont {P.}~\bibnamefont {Mirel}}, \bibinfo {author} {\bibfnamefont
  {S.}~\bibnamefont {Levin}}, \bibinfo {author} {\bibfnamefont
  {M.}~\bibnamefont {Seiffert}}, \ and\ \bibinfo {author} {\bibfnamefont
  {P.~M.}\ \bibnamefont {Lubin}},\ }\href@noop {} {\bibfield  {journal}
  {\bibinfo  {journal} {Rev. Sci. Instrum.}\ }\textbf {\bibinfo
  {volume} {75}},\ \bibinfo {pages} {5079} (\bibinfo {year}
  {2004})}\BibitemShut {NoStop}%
\bibitem [{\citenamefont {Schaeffer}\ \emph {et~al.}(2009)\citenamefont
  {Schaeffer}, \citenamefont {Nucciotti}, \citenamefont {Alessandria},
  \citenamefont {Ardito}, \citenamefont {Barucci}, \citenamefont {Risegari},
  \citenamefont {Ventura}, \citenamefont {Bucci}, \citenamefont {Frossati},
  \citenamefont {Olcese},\ and\ \citenamefont {de~Waard}}]{CUORE}%
  \BibitemOpen
  \bibfield  {author} {\bibinfo {author} {\bibfnamefont {D.}~\bibnamefont
  {Schaeffer}}, \bibinfo {author} {\bibfnamefont {A.}~\bibnamefont
  {Nucciotti}}, \bibinfo {author} {\bibfnamefont {F.}~\bibnamefont
  {Alessandria}}, \bibinfo {author} {\bibfnamefont {R.}~\bibnamefont {Ardito}},
  \bibinfo {author} {\bibfnamefont {M.}~\bibnamefont {Barucci}}, \bibinfo
  {author} {\bibfnamefont {L.}~\bibnamefont {Risegari}}, \bibinfo {author}
  {\bibfnamefont {G.}~\bibnamefont {Ventura}}, \bibinfo {author} {\bibfnamefont
  {C.}~\bibnamefont {Bucci}}, \bibinfo {author} {\bibfnamefont
  {G.}~\bibnamefont {Frossati}}, \bibinfo {author} {\bibfnamefont
  {M.}~\bibnamefont {Olcese}}, \ and\ \bibinfo {author} {\bibfnamefont
  {A.}~\bibnamefont {de~Waard}},\ }\href@noop {} {\bibfield  {journal}
  {\bibinfo  {journal} {Journal of Physics: Conference Series}\ }\textbf
  {\bibinfo {volume} {150}},\ \bibinfo {pages} {012042} (\bibinfo {year}
  {2009})}\BibitemShut {NoStop}%
\bibitem [{\citenamefont {{Xenon 100 Collaboration}}\ \emph
  {et~al.}(2012)\citenamefont {{Xenon 100 Collaboration}}, \citenamefont
  {Aprile} \emph {et~al.}}]{Aprile2012573}%
  \BibitemOpen
  \bibfield  {author} {\bibinfo {author} {\bibnamefont {{Xenon 100
  Collaboration}}}, \bibinfo {author} {\bibfnamefont {E.}~\bibnamefont
  {Aprile}},  \emph {et~al.},\ }\href {\doibase
  http://dx.doi.org/10.1016/j.astropartphys.2012.01.003} {\bibfield  {journal}
  {\bibinfo  {journal} {Astroparticle Physics}\ }\textbf {\bibinfo {volume}
  {35}},\ \bibinfo {pages} {573 } (\bibinfo {year} {2012})}\BibitemShut
  {NoStop}%
\bibitem [{\citenamefont {Storey}\ \emph {et~al.}(2015)\citenamefont {Storey}
  \emph {et~al.}}]{1748-0221-10-02-C02023}%
  \BibitemOpen
  \bibfield  {author} {\bibinfo {author} {\bibfnamefont {J.}~\bibnamefont
  {Storey}} \emph {et~al.},\ }\href
  {http://stacks.iop.org/1748-0221/10/i=02/a=C02023} {\bibfield  {journal}
  {\bibinfo  {journal} {Journal of Instrumentation}\ }\textbf {\bibinfo
  {volume} {10}},\ \bibinfo {pages} {C02023} (\bibinfo {year}
  {2015})}\BibitemShut {NoStop}%
\bibitem [{\citenamefont {Eisaman}\ \emph {et~al.}(2011)\citenamefont
  {Eisaman}, \citenamefont {Fan}, \citenamefont {Migdall},\ and\ \citenamefont
  {Polyakov}}]{singlephoton}%
  \BibitemOpen
  \bibfield  {author} {\bibinfo {author} {\bibfnamefont {M.~D.}\ \bibnamefont
  {Eisaman}}, \bibinfo {author} {\bibfnamefont {J.}~\bibnamefont {Fan}},
  \bibinfo {author} {\bibfnamefont {A.}~\bibnamefont {Migdall}}, \ and\
  \bibinfo {author} {\bibfnamefont {S.~V.}\ \bibnamefont {Polyakov}},\ }\href
  {\doibase http://dx.doi.org/10.1063/1.3610677} {\bibfield  {journal}
  {\bibinfo  {journal} {Rev. Sci. Instrum.}\ }\textbf {\bibinfo
  {volume} {82}},\ \bibinfo {eid} {071101} (\bibinfo {year}
  {2011})}\BibitemShut {NoStop}%
\bibitem [{\citenamefont {Reilly}(2015)}]{NatureQIP}%
  \BibitemOpen
  \bibfield  {author} {\bibinfo {author} {\bibfnamefont {D.~J.}\ \bibnamefont
  {Reilly}},\ }\href@noop {} {\bibfield  {journal} {\bibinfo  {journal} {Nature
 PJ (Quantum Information)}\ }\textbf {\bibinfo {volume} {1}}
  (\bibinfo {year} {2015})}\BibitemShut {NoStop}%
\bibitem [{\citenamefont {Niemack}\ \emph {et~al.}(2008)\citenamefont {Niemack}
  \emph {et~al.}}]{Trans_edge}%
  \BibitemOpen
  \bibfield  {author} {\bibinfo {author} {\bibfnamefont {M.}~\bibnamefont
  {Niemack}} \emph {et~al.},\ }\href {\doibase 10.1007/s10909-008-9729-2}
  {\bibfield  {journal} {\bibinfo  {journal} {J. Low Temp.
  Physics}\ }\textbf {\bibinfo {volume} {151}},\ \bibinfo {pages} {690}
  (\bibinfo {year} {2008})}\BibitemShut {NoStop}%
\bibitem [{\citenamefont {McHugh}\ \emph {et~al.}(2012)\citenamefont {McHugh},
  \citenamefont {Mazin}, \citenamefont {Serfass}, \citenamefont {Meeker},
  \citenamefont {O'Brien}, \citenamefont {Duan}, \citenamefont {Raffanti},\
  and\ \citenamefont {Werthimer}}]{kinetic_induct_MUX}%
  \BibitemOpen
  \bibfield  {author} {\bibinfo {author} {\bibfnamefont {S.}~\bibnamefont
  {McHugh}}, \bibinfo {author} {\bibfnamefont {B.~A.}\ \bibnamefont {Mazin}},
  \bibinfo {author} {\bibfnamefont {B.}~\bibnamefont {Serfass}}, \bibinfo
  {author} {\bibfnamefont {S.}~\bibnamefont {Meeker}}, \bibinfo {author}
  {\bibfnamefont {K.}~\bibnamefont {O'Brien}}, \bibinfo {author} {\bibfnamefont
  {R.}~\bibnamefont {Duan}}, \bibinfo {author} {\bibfnamefont {R.}~\bibnamefont
  {Raffanti}}, \ and\ \bibinfo {author} {\bibfnamefont {D.}~\bibnamefont
  {Werthimer}},\ }\href {\doibase http://dx.doi.org/10.1063/1.3700812}
  {\bibfield  {journal} {\bibinfo  {journal} {Rev. Sci. Instrum.}\ }\textbf {\bibinfo {volume} {83}},\ \bibinfo {eid} {044702}
  (\bibinfo {year} {2012})}\BibitemShut {NoStop}%
\bibitem [{\citenamefont {Colless}\ and\ \citenamefont
  {Reilly}(2014)}]{CollessRSI2}%
  \BibitemOpen
  \bibfield  {author} {\bibinfo {author} {\bibfnamefont {J.~I.}\ \bibnamefont
  {Colless}}\ and\ \bibinfo {author} {\bibfnamefont {D.~J.}\ \bibnamefont
  {Reilly}},\ }\href {\doibase http://dx.doi.org/10.1063/1.4900948} {\bibfield
  {journal} {\bibinfo  {journal} {Rev. Sci. Instrum.}\ }\textbf
  {\bibinfo {volume} {85}},\ \bibinfo {eid} {114706} (\bibinfo {year}
  {2014})}\BibitemShut {NoStop}%
\bibitem [{\citenamefont {Weinreb}\ \emph {et~al.}(2009)\citenamefont
  {Weinreb}, \citenamefont {Bardin}, \citenamefont {Mani},\ and\ \citenamefont
  {Jones}}]{weinreb}%
  \BibitemOpen
  \bibfield  {author} {\bibinfo {author} {\bibfnamefont {S.}~\bibnamefont
  {Weinreb}}, \bibinfo {author} {\bibfnamefont {J.}~\bibnamefont {Bardin}},
  \bibinfo {author} {\bibfnamefont {H.}~\bibnamefont {Mani}}, \ and\ \bibinfo
  {author} {\bibfnamefont {G.}~\bibnamefont {Jones}},\ }\href {\doibase
  http://dx.doi.org/10.1063/1.3103939} {\bibfield  {journal} {\bibinfo
  {journal} {Rev. Sci. Instrum.}\ }\textbf {\bibinfo {volume}
  {80}},\ \bibinfo {eid} {044702} (\bibinfo {year} {2009})}\BibitemShut
  {NoStop}%
\bibitem [{\citenamefont {Hornibrook}\ \emph {et~al.}(2014)\citenamefont
  {Hornibrook}, \citenamefont {Colless}, \citenamefont {Mahoney}, \citenamefont
  {Croot}, \citenamefont {Blanvillain}, \citenamefont {Lu}, \citenamefont
  {Gossard},\ and\ \citenamefont {Reilly}}]{Hornibrook_APL}%
  \BibitemOpen
  \bibfield  {author} {\bibinfo {author} {\bibfnamefont {J.~M.}\ \bibnamefont
  {Hornibrook}}, \bibinfo {author} {\bibfnamefont {J.~I.}\ \bibnamefont
  {Colless}}, \bibinfo {author} {\bibfnamefont {A.~C.}\ \bibnamefont
  {Mahoney}}, \bibinfo {author} {\bibfnamefont {X.~G.}\ \bibnamefont {Croot}},
  \bibinfo {author} {\bibfnamefont {S.}~\bibnamefont {Blanvillain}}, \bibinfo
  {author} {\bibfnamefont {H.}~\bibnamefont {Lu}}, \bibinfo {author}
  {\bibfnamefont {A.~C.}\ \bibnamefont {Gossard}}, \ and\ \bibinfo {author}
  {\bibfnamefont {D.~J.}\ \bibnamefont {Reilly}},\ }\href@noop {} {\bibfield
  {journal} {\bibinfo  {journal} {App. Phys. Lett.}\ }\textbf {\bibinfo
  {volume} {104}},\ \bibinfo {pages} {103108} (\bibinfo {year}
  {2014})}\BibitemShut {NoStop}%
\bibitem [{\citenamefont {Al-Taie}\ \emph {et~al.}(2015)\citenamefont
  {Al-Taie}, \citenamefont {Smith}, \citenamefont {Lesage}, \citenamefont
  {See}, \citenamefont {Griffiths}, \citenamefont {Beere}, \citenamefont
  {Jones}, \citenamefont {Ritchie}, \citenamefont {Kelly},\ and\ \citenamefont
  {Smith}}]{Smith_mux}%
  \BibitemOpen
  \bibfield  {author} {\bibinfo {author} {\bibfnamefont {H.}~\bibnamefont
  {Al-Taie}}, \bibinfo {author} {\bibfnamefont {L.~W.}\ \bibnamefont {Smith}},
  \bibinfo {author} {\bibfnamefont {A.~A.~J.}\ \bibnamefont {Lesage}}, \bibinfo
  {author} {\bibfnamefont {P.}~\bibnamefont {See}}, \bibinfo {author}
  {\bibfnamefont {J.~P.}\ \bibnamefont {Griffiths}}, \bibinfo {author}
  {\bibfnamefont {H.~E.}\ \bibnamefont {Beere}}, \bibinfo {author}
  {\bibfnamefont {G.~A.~C.}\ \bibnamefont {Jones}}, \bibinfo {author}
  {\bibfnamefont {D.~A.}\ \bibnamefont {Ritchie}}, \bibinfo {author}
  {\bibfnamefont {M.~J.}\ \bibnamefont {Kelly}}, \ and\ \bibinfo {author}
  {\bibfnamefont {C.~G.}\ \bibnamefont {Smith}},\ }\href {\doibase
  http://dx.doi.org/10.1063/1.4928615} {\bibfield  {journal} {\bibinfo
  {journal} {J. App. Phys.}\ }\textbf {\bibinfo {volume} {118}},\
  \bibinfo {eid} {075703} (\bibinfo {year} {2015})}\BibitemShut {NoStop}%
\bibitem [{\citenamefont {Colless}\ \emph {et~al.}(2013)\citenamefont
  {Colless}, \citenamefont {Mahoney}, \citenamefont {Hornibrook}, \citenamefont
  {Doherty}, \citenamefont {Lu}, \citenamefont {Gossard},\ and\ \citenamefont
  {Reilly}}]{Colless_PRL}%
  \BibitemOpen
  \bibfield  {author} {\bibinfo {author} {\bibfnamefont {J.~I.}\ \bibnamefont
  {Colless}}, \bibinfo {author} {\bibfnamefont {A.~C.}\ \bibnamefont
  {Mahoney}}, \bibinfo {author} {\bibfnamefont {J.~M.}\ \bibnamefont
  {Hornibrook}}, \bibinfo {author} {\bibfnamefont {A.~C.}\ \bibnamefont
  {Doherty}}, \bibinfo {author} {\bibfnamefont {H.}~\bibnamefont {Lu}},
  \bibinfo {author} {\bibfnamefont {A.~C.}\ \bibnamefont {Gossard}}, \ and\
  \bibinfo {author} {\bibfnamefont {D.~J.}\ \bibnamefont {Reilly}},\
  }\href@noop {} {\bibfield  {journal} {\bibinfo  {journal} {Phys. Rev. Lett}\
  }\textbf {\bibinfo {volume} {110}},\ \bibinfo {pages} {046805} (\bibinfo
  {year} {2013})}\BibitemShut {NoStop}%
\bibitem [{\citenamefont {Rahman}\ and\ \citenamefont
  {Lehmann}(2014)}]{6908339}%
  \BibitemOpen
  \bibfield  {author} {\bibinfo {author} {\bibfnamefont {M.}~\bibnamefont
  {Rahman}}\ and\ \bibinfo {author} {\bibfnamefont {T.}~\bibnamefont
  {Lehmann}},\ }in\ \href {\doibase 10.1109/MWSCAS.2014.6908339} {\emph
  {\bibinfo {booktitle} {Circuits and Systems (MWSCAS), 2014 IEEE 57th
  International Midwest Symposium on}}}\ (\bibinfo {year} {2014})\ pp.\
  \bibinfo {pages} {9--12}\BibitemShut {NoStop}%
\bibitem [{\citenamefont {Takahashi}\ \emph {et~al.}(2014)\citenamefont
  {Takahashi}, \citenamefont {Shimada}, \citenamefont {Maezawa},\ and\
  \citenamefont {Mizugaki}}]{Takahashi2014220}%
  \BibitemOpen
  \bibfield  {author} {\bibinfo {author} {\bibfnamefont {Y.}~\bibnamefont
  {Takahashi}}, \bibinfo {author} {\bibfnamefont {H.}~\bibnamefont {Shimada}},
  \bibinfo {author} {\bibfnamefont {M.}~\bibnamefont {Maezawa}}, \ and\
  \bibinfo {author} {\bibfnamefont {Y.}~\bibnamefont {Mizugaki}},\ }\href
  {\doibase http://dx.doi.org/10.1016/j.phpro.2014.09.060} {\bibfield
  {journal} {\bibinfo  {journal} {Physics Procedia}\ }\textbf {\bibinfo
  {volume} {58}},\ \bibinfo {pages} {220} (\bibinfo {year} {2014})},\ \bibinfo
  {note} {proceedings of the 26th International Symposium on Superconductivity
  (ISS 2013)}\BibitemShut {NoStop}%
\bibitem [{\citenamefont {Mukhanov}(2011)}]{SFQ_ADC}%
  \BibitemOpen
  \bibfield  {author} {\bibinfo {author} {\bibfnamefont {O.~A.}\ \bibnamefont
  {Mukhanov}},\ }\href@noop {} {\emph {\bibinfo {title} {History of
  Superconductor Analog-to-Digital Converters, in 100 Years of Superconductivity, H. Rogalla and P. Kes, Ed. Taylor \& Francis, London, UK}}}\ (\bibinfo {year}
  {2011})\BibitemShut {NoStop}%
\bibitem [{\citenamefont {Okcan}\ \emph {et~al.}(2010)\citenamefont {Okcan},
  \citenamefont {Merken}, \citenamefont {Gielen},\ and\ \citenamefont
  {Van~Hoof}}]{cryoADC}%
  \BibitemOpen
  \bibfield  {author} {\bibinfo {author} {\bibfnamefont {B.}~\bibnamefont
  {Okcan}}, \bibinfo {author} {\bibfnamefont {P.}~\bibnamefont {Merken}},
  \bibinfo {author} {\bibfnamefont {G.}~\bibnamefont {Gielen}}, \ and\ \bibinfo
  {author} {\bibfnamefont {C.}~\bibnamefont {Van~Hoof}},\ }\href {\doibase
  http://dx.doi.org/10.1063/1.3309825} {\bibfield  {journal} {\bibinfo
  {journal} {Rev. Sci. Instrum.}\ }\textbf {\bibinfo {volume}
  {81}},\ \bibinfo {eid} {024702} (\bibinfo {year} {2010})}\BibitemShut
  {NoStop}%
\bibitem [{\citenamefont {Tighe}\ \emph {et~al.}(1999)\citenamefont {Tighe},
  \citenamefont {Akerling},\ and\ \citenamefont {Smith}}]{MCM_ribbonSC}%
  \BibitemOpen
  \bibfield  {author} {\bibinfo {author} {\bibfnamefont {T.~S.}\ \bibnamefont
  {Tighe}}, \bibinfo {author} {\bibfnamefont {G.}~\bibnamefont {Akerling}}, \
  and\ \bibinfo {author} {\bibfnamefont {A.~D.}\ \bibnamefont {Smith}},\
  }\href@noop {} {\bibfield  {journal} {\bibinfo  {journal} {IEEE Trans.
  Applied Superconductivity}\ }\textbf {\bibinfo {volume} {9}},\ \bibinfo
  {pages} {3173} (\bibinfo {year} {1999})}\BibitemShut {NoStop}%
\bibitem [{\citenamefont {Yung}\ and\ \citenamefont {Moeckly}(2011)}]{5613217}%
  \BibitemOpen
  \bibfield  {author} {\bibinfo {author} {\bibfnamefont {C.}~\bibnamefont
  {Yung}}\ and\ \bibinfo {author} {\bibfnamefont {B.}~\bibnamefont {Moeckly}},\
  }\href {\doibase 10.1109/TASC.2010.2080655} {\bibfield  {journal} {\bibinfo
  {journal} {IEEE Trans. Applied Superconductivity}\ }\textbf
  {\bibinfo {volume} {21}},\ \bibinfo {pages} {107} (\bibinfo {year}
  {2011})}\BibitemShut {NoStop}%
\bibitem [{\citenamefont {Hornibrook}\ \emph {et~al.}(2015)\citenamefont
  {Hornibrook}, \citenamefont {Colless}, \citenamefont {Conway~Lamb},
  \citenamefont {Pauka}, \citenamefont {Lu}, \citenamefont {Gossard},
  \citenamefont {Watson}, \citenamefont {Gardner}, \citenamefont {Fallahi},
  \citenamefont {Manfra},\ and\ \citenamefont
  {Reilly}}]{PhysRevApplied.3.024010}%
  \BibitemOpen
  \bibfield  {author} {\bibinfo {author} {\bibfnamefont {J.~M.}\ \bibnamefont
  {Hornibrook}}, \bibinfo {author} {\bibfnamefont {J.~I.}\ \bibnamefont
  {Colless}}, \bibinfo {author} {\bibfnamefont {I.~D.}\ \bibnamefont
  {Conway~Lamb}}, \bibinfo {author} {\bibfnamefont {S.~J.}\ \bibnamefont
  {Pauka}}, \bibinfo {author} {\bibfnamefont {H.}~\bibnamefont {Lu}}, \bibinfo
  {author} {\bibfnamefont {A.~C.}\ \bibnamefont {Gossard}}, \bibinfo {author}
  {\bibfnamefont {J.~D.}\ \bibnamefont {Watson}}, \bibinfo {author}
  {\bibfnamefont {G.~C.}\ \bibnamefont {Gardner}}, \bibinfo {author}
  {\bibfnamefont {S.}~\bibnamefont {Fallahi}}, \bibinfo {author} {\bibfnamefont
  {M.~J.}\ \bibnamefont {Manfra}}, \ and\ \bibinfo {author} {\bibfnamefont
  {D.~J.}\ \bibnamefont {Reilly}},\ }\href {\doibase
  10.1103/PhysRevApplied.3.024010} {\bibfield  {journal} {\bibinfo  {journal}
  {Phys. Rev. Applied}\ }\textbf {\bibinfo {volume} {3}},\ \bibinfo {pages}
  {024010} (\bibinfo {year} {2015})}\BibitemShut {NoStop}%
\bibitem [{\citenamefont {Shulman}\ \emph {et~al.}(2014)\citenamefont
  {Shulman}, \citenamefont {Harvey}, \citenamefont {Nichol}, \citenamefont
  {Bartlett}, \citenamefont {Doherty}, \citenamefont {Umansky},\ and\
  \citenamefont {Yacoby}}]{Yacobyfeedback}%
  \BibitemOpen
  \bibfield  {author} {\bibinfo {author} {\bibfnamefont {M.~D.}\ \bibnamefont
  {Shulman}}, \bibinfo {author} {\bibfnamefont {S.~P.}\ \bibnamefont {Harvey}},
  \bibinfo {author} {\bibfnamefont {J.~M.}\ \bibnamefont {Nichol}}, \bibinfo
  {author} {\bibfnamefont {S.~D.}\ \bibnamefont {Bartlett}}, \bibinfo {author}
  {\bibfnamefont {A.~C.}\ \bibnamefont {Doherty}}, \bibinfo {author}
  {\bibfnamefont {V.}~\bibnamefont {Umansky}}, \ and\ \bibinfo {author}
  {\bibfnamefont {A.}~\bibnamefont {Yacoby}},\ }\href@noop {} {\bibfield
  {journal} {\bibinfo  {journal} {Nature Comm.}\ }\textbf {\bibinfo {volume}
  {5}} (\bibinfo {year} {2014})}\BibitemShut {NoStop}%
\bibitem [{\citenamefont {Rist\`{e}}\ and\ \citenamefont
  {DiCarlo}(2015)}]{Riste}%
  \BibitemOpen
  \bibfield  {author} {\bibinfo {author} {\bibfnamefont {D.}~\bibnamefont
  {Rist\`{e}}}\ and\ \bibinfo {author} {\bibfnamefont {L.}~\bibnamefont
  {DiCarlo}},\ }\href@noop {} {\bibfield  {journal} {\bibinfo  {journal}
  {arXiv:1508.01385}\ } (\bibinfo {year} {2015})}\BibitemShut {NoStop}%
\bibitem [{\citenamefont {{Xilinx White Paper}}(2013)}]{Xilinx:28nmHf}%
  \BibitemOpen
  \bibfield  {author} {\bibinfo {author} {\bibnamefont {{Xilinx White
  Paper}}},\ }\href
  {http://www.xilinx.com/support/documentation/white_papers/wp312_Next_Gen_28_nm_Overview.pdf}
  {\enquote {\bibinfo {title} {{Xilinx Next Generation 28~nm FPGA Technology
  Overview v1.1.1}},}\ }\bibinfo {howpublished}
  {\url{http://www.xilinx.com/support/documentation/white_papers/wp312_Next_Gen_28_nm_Overview.pdf}}
  (\bibinfo {year} {2013})\BibitemShut {NoStop}%
\bibitem [{\citenamefont {{Xilinx Selection
  Guide}}(2015)}]{Xilinx:A7ProductTable}%
  \BibitemOpen
  \bibfield  {author} {\bibinfo {author} {\bibnamefont {{Xilinx Selection
  Guide}}},\ }\href
  {http://www.xilinx.com/support/documentation/selection-guides/artix7-product-table.pdf}
  {\enquote {\bibinfo {title} {{Artix-7 FPGAs v4.7}},}\ }\bibinfo
  {howpublished}
  {\url{http://www.xilinx.com/support/documentation/selection-guides/artix7-product-table.pdf}}
  (\bibinfo {year} {2015})\BibitemShut {NoStop}%
\bibitem [{\citenamefont {{Xilinx User
  Guide}}(2014{\natexlab{a}})}]{Xilinx:DSP48E1}%
  \BibitemOpen
  \bibfield  {author} {\bibinfo {author} {\bibnamefont {{Xilinx User Guide}}},\
  }\href
  {http://www.xilinx.com/support/documentation/user_guides/ug479_7Series_DSP48E1.pdf}
  {\enquote {\bibinfo {title} {{7 Series DSP48E1 Slice v1.8}},}\ }\bibinfo
  {howpublished}
  {\url{http://www.xilinx.com/support/documentation/user_guides/ug479_7Series_DSP48E1.pdf}}
  (\bibinfo {year} {2014}{\natexlab{a}})\BibitemShut {NoStop}%
\bibitem [{\citenamefont {{Xilinx Datasheet}}(2015)}]{Xilinx:Switching}%
  \BibitemOpen
  \bibfield  {author} {\bibinfo {author} {\bibnamefont {{Xilinx Datasheet}}},\
  }\href
  {http://www.xilinx.com/support/documentation/data_sheets/ds181_Artix_7_Data_Sheet.pdf}
  {\enquote {\bibinfo {title} {{Artix-7 FPGAs Data Sheet: DC and AC Switching
  Characteristics v1.18}},}\ }\bibinfo {howpublished}
  {\url{http://www.xilinx.com/support/documentation/data_sheets/ds181_Artix_7_Data_Sheet.pdf}}
  (\bibinfo {year} {2015})\BibitemShut {NoStop}%
\bibitem [{\citenamefont {Teyssandier}\ and\ \citenamefont
  {Pr{\^e}le}(2010)}]{Teyssandier:2010}%
  \BibitemOpen
  \bibfield  {author} {\bibinfo {author} {\bibfnamefont {F.}~\bibnamefont
  {Teyssandier}}\ and\ \bibinfo {author} {\bibfnamefont {D.}~\bibnamefont
  {Pr{\^e}le}},\ }in\ \href {http://hal.archives-ouvertes.fr/hal-00623399}
  {\emph {\bibinfo {booktitle} {{Ninth International Workshop on Low
  Temperature Electronics - WOLTE9}}}}\ (\bibinfo {year} {2010})\BibitemShut
  {NoStop}%
\bibitem [{\citenamefont {Teverovsky}(2005)}]{NASA:Capacitors}%
  \BibitemOpen
  \bibfield  {author} {\bibinfo {author} {\bibfnamefont {A.}~\bibnamefont
  {Teverovsky}},\ }\href@noop {} {\enquote {\bibinfo {title} {{Reliability of
  Electronics at Cryogenic Temperatures}},}\ }\bibinfo {howpublished}
  {\url{http://nepp.nasa.gov/DocUploads/1BCC6326-46F7-4C5E-94D476A8B6176ED1/Approach\%20to\%20Reliability\%20Testing\%20at\%20cryo\%20NEPP.pdf}}
  (\bibinfo {year} {2005}),\ \bibinfo {note} {[Online; accessed
  02-Jul-2013]}\BibitemShut {NoStop}%
\bibitem [{\citenamefont {{Xilinx User
  Guide}}(2014{\natexlab{b}})}]{Xilinx:7CLB}%
  \BibitemOpen
  \bibfield  {author} {\bibinfo {author} {\bibnamefont {{Xilinx User Guide}}},\
  }\href
  {http://www.xilinx.com/support/documentation/user_guides/ug474_7Series_CLB.pdf}
  {\enquote {\bibinfo {title} {{7-series Configurable Logic Block v1.7}},}\
  }\bibinfo {howpublished}
  {\url{http://www.xilinx.com/support/documentation/user_guides/ug474_7Series_CLB.pdf}}
  (\bibinfo {year} {2014}{\natexlab{b}})\BibitemShut {NoStop}%
\bibitem [{\citenamefont {{Xilinx Product Guide}}(2014)}]{Xilinx:Microblaze}%
  \BibitemOpen
  \bibfield  {author} {\bibinfo {author} {\bibnamefont {{Xilinx Product
  Guide}}},\ }\href
  {http://www.xilinx.com/support/documentation/sw_manuals/xilinx2014_2/pg116-microblaze-mcs.pdf}
  {\enquote {\bibinfo {title} {{LogiCORE IP MicroBlaze Micro Controller System
  v2.2}},}\ }\bibinfo {howpublished}
  {\url{http://www.xilinx.com/support/documentation/sw_manuals/xilinx2014_2/pg116-microblaze-mcs.pdf}}
  (\bibinfo {year} {2014})\BibitemShut {NoStop}%
\bibitem [{\citenamefont {{ARM Ltd.}}(2008)}]{ARM_M1}%
  \BibitemOpen
  \bibfield  {author} {\bibinfo {author} {\bibnamefont {{ARM Ltd.}}},\ }\href
  {http://infocenter.arm.com/help/topic/com.arm.doc.ddi0413d/DDI0413D_cortexm1_r1p0_trm.pdf}
  {\enquote {\bibinfo {title} {{Cortex-M1 Technical Reference Manual Issue
  D}},}\ }\bibinfo {howpublished}
  {\url{http://infocenter.arm.com/help/topic/com.arm.doc.ddi0413d/DDI0413D_cortexm1_r1p0_trm.pdf}}
  (\bibinfo {year} {2008})\BibitemShut {NoStop}%
\bibitem [{\citenamefont {Weinreb}\ \emph {et~al.}(2007)\citenamefont
  {Weinreb}, \citenamefont {Bardin},\ and\ \citenamefont {Mani}}]{CryoSiGeAmp}%
  \BibitemOpen
  \bibfield  {author} {\bibinfo {author} {\bibfnamefont {S.}~\bibnamefont
  {Weinreb}}, \bibinfo {author} {\bibfnamefont {J.}~\bibnamefont {Bardin}}, \
  and\ \bibinfo {author} {\bibfnamefont {H.}~\bibnamefont {Mani}},\ }\href
  {\doibase 10.1109/TMTT.2007.907729} {\bibfield  {journal} {\bibinfo
  {journal} {IEEE Trans. Microwave Theory and Techniques}\ }\textbf
  {\bibinfo {volume} {55}},\ \bibinfo {pages} {2306} (\bibinfo {year}
  {2007})}\BibitemShut {NoStop}%
\bibitem [{\citenamefont {Cressler}(2010)}]{Cressler:2010}%
  \BibitemOpen
  \bibfield  {author} {\bibinfo {author} {\bibfnamefont {J.}~\bibnamefont
  {Cressler}},\ }\href {\doibase 10.1109/TDMR.2010.2050691} {\bibfield
  {journal} {\bibinfo  {journal} {IEEE
  Trans. Device and Materials Reliability, }\ }\textbf {\bibinfo {volume} {10}},\ \bibinfo {pages} {437}
  (\bibinfo {year} {2010})}\BibitemShut {NoStop}%
\bibitem [{\citenamefont {Hanamura}\ \emph {et~al.}(1986)\citenamefont
  {Hanamura}, \citenamefont {Aoki}, \citenamefont {Masuhara}, \citenamefont
  {Minato}, \citenamefont {Sakai},\ and\ \citenamefont
  {Hayashida}}]{Hanamura:1986}%
  \BibitemOpen
  \bibfield  {author} {\bibinfo {author} {\bibfnamefont {S.}~\bibnamefont
  {Hanamura}}, \bibinfo {author} {\bibfnamefont {M.}~\bibnamefont {Aoki}},
  \bibinfo {author} {\bibfnamefont {T.}~\bibnamefont {Masuhara}}, \bibinfo
  {author} {\bibfnamefont {O.}~\bibnamefont {Minato}}, \bibinfo {author}
  {\bibfnamefont {Y.}~\bibnamefont {Sakai}}, \ and\ \bibinfo {author}
  {\bibfnamefont {T.}~\bibnamefont {Hayashida}},\ }\href {\doibase
  10.1109/JSSC.1986.1052555} {\bibfield  {journal} {\bibinfo  {journal}
  {IEEE J. Solid-State Circuits}\ }\textbf {\bibinfo {volume} {21}},\
  \bibinfo {pages} {484} (\bibinfo {year} {1986})}\BibitemShut {NoStop}%
\bibitem [{\citenamefont {Clark}\ \emph {et~al.}(1992)\citenamefont {Clark},
  \citenamefont {El-Kareh}, \citenamefont {Pires}, \citenamefont {Titcomb},\
  and\ \citenamefont {Anderson}}]{Clark:1992}%
  \BibitemOpen
  \bibfield  {author} {\bibinfo {author} {\bibfnamefont {W.}~\bibnamefont
  {Clark}}, \bibinfo {author} {\bibfnamefont {B.}~\bibnamefont {El-Kareh}},
  \bibinfo {author} {\bibfnamefont {R.}~\bibnamefont {Pires}}, \bibinfo
  {author} {\bibfnamefont {S.}~\bibnamefont {Titcomb}}, \ and\ \bibinfo
  {author} {\bibfnamefont {R.}~\bibnamefont {Anderson}},\ }\href {\doibase
  10.1109/33.148509} {\bibfield  {journal} {\bibinfo  {journal} {IEEE Trans. Components,
  Hybrids, and Manufacturing Technology}\ }\textbf
  {\bibinfo {volume} {15}},\ \bibinfo {pages} {397} (\bibinfo {year}
  {1992})}\BibitemShut {NoStop}%
\bibitem [{\citenamefont {Kuon}\ and\ \citenamefont {Rose}(2007)}]{FPGAvsASIC}%
  \BibitemOpen
  \bibfield  {author} {\bibinfo {author} {\bibfnamefont {I.}~\bibnamefont
  {Kuon}}\ and\ \bibinfo {author} {\bibfnamefont {J.}~\bibnamefont {Rose}},\
  }\href {\doibase 10.1109/TCAD.2006.884574} {\bibfield  {journal} {\bibinfo
  {journal} {IEEE Trans. Computer-Aided Design of Integrated Circuits and Systems}\ }\textbf {\bibinfo {volume} {26}},\ \bibinfo {pages} {203}
  (\bibinfo {year} {2007})}\BibitemShut {NoStop}%
\bibitem [{\citenamefont {Likharev}\ and\ \citenamefont
  {Semenov}(1991)}]{likharev1991rsfq}%
  \BibitemOpen
  \bibfield  {author} {\bibinfo {author} {\bibfnamefont {K.~K.}\ \bibnamefont
  {Likharev}}\ and\ \bibinfo {author} {\bibfnamefont {V.~K.}\ \bibnamefont
  {Semenov}},\ }\href@noop {} {\bibfield  {journal} {\bibinfo  {journal}
  {IEEE Trans. Applied Superconductivity}\ }\textbf {\bibinfo
  {volume} {1}},\ \bibinfo {pages} {3} (\bibinfo {year} {1991})}\BibitemShut
  {NoStop}%
\bibitem [{\citenamefont {Herr}\ \emph {et~al.}(2011)\citenamefont {Herr},
  \citenamefont {Herr}, \citenamefont {Oberg},\ and\ \citenamefont
  {Ioannidis}}]{rql}%
  \BibitemOpen
  \bibfield  {author} {\bibinfo {author} {\bibfnamefont {Q.~P.}\ \bibnamefont
  {Herr}}, \bibinfo {author} {\bibfnamefont {A.~Y.}\ \bibnamefont {Herr}},
  \bibinfo {author} {\bibfnamefont {O.~T.}\ \bibnamefont {Oberg}}, \ and\
  \bibinfo {author} {\bibfnamefont {A.~G.}\ \bibnamefont {Ioannidis}},\ }\href
  {\doibase http://dx.doi.org/10.1063/1.3585849} {\bibfield  {journal}
  {\bibinfo  {journal} {J. App. Physics}\ }\textbf {\bibinfo
  {volume} {109}},\ \bibinfo {eid} {103903} (\bibinfo {year}
  {2011})}\BibitemShut {NoStop}%
\bibitem [{\citenamefont {Vernik}\ \emph {et~al.}(2007)\citenamefont {Vernik},
  \citenamefont {Kirichenko}, \citenamefont {Dotsenko}, \citenamefont {Miller},
  \citenamefont {Webber}, \citenamefont {Shevchenko}, \citenamefont
  {Talalaevskii}, \citenamefont {Gupta},\ and\ \citenamefont
  {Mukhanov}}]{vernik2007cryocooled}%
  \BibitemOpen
  \bibfield  {author} {\bibinfo {author} {\bibfnamefont {I.~V.}\ \bibnamefont
  {Vernik}}, \bibinfo {author} {\bibfnamefont {D.~E.}\ \bibnamefont
  {Kirichenko}}, \bibinfo {author} {\bibfnamefont {V.~V.}\ \bibnamefont
  {Dotsenko}}, \bibinfo {author} {\bibfnamefont {R.}~\bibnamefont {Miller}},
  \bibinfo {author} {\bibfnamefont {R.~J.}\ \bibnamefont {Webber}}, \bibinfo
  {author} {\bibfnamefont {P.}~\bibnamefont {Shevchenko}}, \bibinfo {author}
  {\bibfnamefont {A.}~\bibnamefont {Talalaevskii}}, \bibinfo {author}
  {\bibfnamefont {D.}~\bibnamefont {Gupta}}, \ and\ \bibinfo {author}
  {\bibfnamefont {O.~A.}\ \bibnamefont {Mukhanov}},\ }\href@noop {} {\bibfield
  {journal} {\bibinfo  {journal} {Superconductor Science and Technology}\
  }\textbf {\bibinfo {volume} {20}},\ \bibinfo {pages} {S323} (\bibinfo {year}
  {2007})}\BibitemShut {NoStop}%
\end{thebibliography}
%

\end{document}